# Generative Knowledge Production Pipeline Driven by Academic Influencers

Katalin FEHER
feher.katalin@uni-nke.hu
Ludovika University of Public Service

Marton DEMETER
demeter.marton@uni-nke.hu
Ludovika University of Public Service

Horizon Europe NGI, Grant No. 101070125

**ABSTRACT**

Generative AI transforms knowledge production, validation, and dissemination, raising academic integrity and credibility concerns. This study examines 53 academic influencer videos that reached 5.3 million viewers to identify an emerging, structured, implementation-ready pipeline balancing originality, ethical compliance, and human-AI collaboration despite the disruptive impacts. Findings highlight generative AI's potential to automate publication workflows and democratize participation in knowledge production while challenging traditional scientific norms. Academic influencers emerge as key intermediaries in this paradigm shift, connecting bottom-up practices with institutional policies to improve adaptability. Accordingly, the study proposes a generative publication production pipeline and a policy framework for co-intelligence adaptation and reinforcing credibility-centered standards in AI-powered research. These insights support scholars, educators, and policymakers in understanding AI's transformative impact by advocating responsible and innovation-driven knowledge production. Additionally, they reveal pathways for automating best practices, optimizing scholarly workflows, and fostering creativity in academic research and publication.

*Keywords: generative AI, ChatPGT, academic integrity, influencers, knowledge production, social media, policy implications, academic policy*

## 1. INTRODUCTION

The advent of generative AI (GenAI) transforms knowledge production, increasingly supporting and partially automating the academic workflow (Bolanos et al. 2024). This trend suggests a paradigm shift where researchers utilize effectively and productively generative AI tools, potentially leading to more automated scientific workflows.

However, we have also identified a human component in this process: the impact of the academic influencers via social media promoting hands-on knowledge about GenAI in academic projects. With our extensive expertise as researchers, mentors, and experts across universities and academic cultures spanning multiple continents, we have observed how ambitious researchers are trying to keep up with the technological pace and gain a competitive advantage from the features of emerging generative AI tools. These scholars and students are exploring the evolving academic landscape by turning to video influencers who provide precise, practical guidance, often filling the gaps left by traditional policies. Generative AI represents a transformative catalyst, reshaping industries and academia alike—thus, we have sought to capture this transformative moment in the history of academic knowledge production. Such grassroots innovation, supported by generative AI, underscores the urgency of studying how these forces transform knowledge production, fostering a more adaptable, inclusive academic ecosystem via globally available narratives.

This study investigates how academic influencers promote effective, productive and ethical publication automation workflow leveraging generative AI, particularly ChatGPT. Analyzing how these influencers establish influential practices and norms, we identify a structured pipeline that enables the automation of scholarly processes while balancing originality, credibility, and institutional compliance. This framework reveals how grassroots practices intersect with policy-driven governance, raising questions about whether AI-driven research marks a paradigm shift or destabilizes academic integrity. Existing research mainly focuses on institutions' and prominent publishers' top-down generative AI policies; thus, they have overlooked the structured mechanisms by which these influencers enable automation and also advocate academic ethical standards and credibility. This study addresses that void by critically assessing whether their role fosters responsible AI adoption or instead accelerates an informal, unregulated shift in academic integrity. Our findings provide a necessary intervention by revealing the hidden influence of academic influencers and proposing a pipeline structure that explicitly integrates ethical compliance—ensuring that AI-driven scholarship remains accountable rather than dictated by viral, unchecked trends.

## 2. GENERATIVE TECHNOLOGY IS CHALLENGING ACADEMIC INTEGRITY

Generative AI, and ChatGPT in particular, has rapidly become an essential tool in academic publishing for analyzing unstructured data, and generating human-like texts (Dwivedi et al., 2023), promising greater efficiency and improved research quality. As these tools reshape content creation (Feher 2025), writing and working style (Merrill & Lerman 2024) and literature reviews, the evolution of academic writing standards raises ethical concerns about plagiarism, biases, and research credibility (Nam & Bai, 2023; Livberber & Ayvaz, 2023; Sun et al., 2023, Geraldi 2021). Balancing AI's benefits against rigorous academic standards has, therefore, presented a complex challenge since ChatGPT first appeared on the scene. This challenge has sparked extensive debates and a growing body of literature examining its implications for academic integrity.

ChatGPT has become a cornerstone of modern academic workflows, dominating video tutorials on generative AI due to its accessibility and broad functionality. Advances in AI bring domain-specific and multi-modal models such as DARWIN19, Bloom20, and KOSMOS21 into the spotlight, which integrate language and visuals with tools like Scholar AI that are tailored for academia. Despite these innovations, ChatGPT remains the preferred tool among scholars due to its accessibility, cost-effectiveness, versatility, ease of integration, and widespread student adoption, which are shaping research practices and academic workflows globally.

The technological evolutionary leap through generative AI (Feher 2025) has fundamentally transformed academic knowledge production and norms, one symbolic example of which is the way ChatGPT facilitates AI socialization across all levels of study, research, and publication. Its widespread adoption has significantly improved productivity by bridging gaps between traditional academic practices and AI-driven capabilities. As scholars and students adapt to its integration, ChatGPT continues to redefine research methodologies, challenge established standards of academic integrity, and compel academia to evolve in response to this transformative technological shift.

Top-down policies from universities and publishers (Gering et al. 2025, Gulumbe, 2024) dominate discussions on academic integrity in this era of generative AI, echoing earlier knowledge production frameworks and their ethics and credibility criteria. Publishers such as Elsevier and Sage have adopted COPE guidelines (Cacciamani et al., 2023) that exclude AI from authorship and require human accountability for all content (Demeter & Halo, 2025). This conservative approach, however, may clash with the bottom-up adaptation seen among young scholars and students, who rapidly integrate generative tools into their workflows—raising questions about evolving norms, institutional flexibility, and the future of credible knowledge-making.

At the same time, bottom-up influences, led by ChatGPT video influencers, are transforming research and publication practices. Unlike historical Luddism, which resisted technological change (Darici, 2024), these influencers actively promote generative AI tools, emphasizing their competitive advantage for scholars. While Lim (2023) and Sun et al. (2023) have examined the paradox of AI as both a banned and popular technology, along with the ethical challenges it poses (Toh & Park 2025, Gregoire et al. 2024, Tomasev et al. 2020), the impact of academic influencers in shaping informal norms remains largely unexplored.

ChatGPT influencers leverage AI to improve productivity and innovation, democratizing access to knowledge while monetizing their platforms. This individual-driven approach offers opportunities but also fragments academic standards, as financial priorities often overshadow rigorous practices. While influencers expand AI's accessibility and inspire new research areas, concerns about plagiarism and misinformation persist (Rasul et al. 2023; Livberber & Ayvaz, 2023). A realistic concept of academic integrity requires resilience and attention to bottom-up approaches because transformative technologies and evolving academic trends challenge traditional norms.

To sum up, generative AI transforms research and publication by streamlining and redefining tasks (Dwivedi et al. 2024), signifying a paradigm shift in academia (Kashyap, 2024). Its growing influence necessitates an examination of its effects on research productivity and academic integrity. This study explores the evolving practices of publication and academic knowledge generation promoted by ChatGPT video influencers via social media, exploring the broader implications of paradigm shifts in academia and offering insights to guide related academic policies.

## 3. GENERATIVE AND INFLUENCER-DRIVEN PARADIGM SHIFT

The paradigm shifts in academia, from the transition to written traditions to the rise of generative AI, have significantly transformed how knowledge is produced, validated, and disseminated (see Table 1). The fundamental paradigm shifts in academia currently show how evolving human, digital, and AI-driven practices transform knowledge production and challenge academic integrity, credibility and authorial authenticity (Watson et al. 2025). These shifts have expanded democratized knowledge (Tayebi et al. 2024), challenged equity, and redefined academic trust (Demeter 2020). Recent social media and AI advances are profoundly transforming academia through synthetic content (Feher 2025), building on the earlier digital transformation movement that championed the democratization of knowledge. The next paradigm shift may see AI as the primary driver of research, applying human-machine co-intelligence and communication (Mollick, 2024; Hoffmann et al., 2024, Feher et al. 2024, Feher & Katona 2021), thus redefining intellectual property. Recent advances in AI have sparked ethical debates, prompting guidelines to address AI's disruptive potential and to establish normative principles (Hagendorff, 2020). The philosophical foundation of academia, therefore, faces scrutiny as algorithmic ghostwriting and democratized creation challenge originality, authorship, and the ethical essence of scholarly contributions.

*Table 1. Academic Paradigm Shifts and their Impacts. Source: The Authors*

| Paradigm Shifts | Socio-technical factors | Key Developments | Social Impacts |
|---|---|---|---|
| **Oral to written traditions** | Human | Formal attribution | Need for clear attribution, increase in plagiarism concerns |
| **Scientific method emergence** | Human | Empiricism and transparency | Increased credibility and reliability in scientific research |
| **Peer review institutionalization** | Human | Quality assurance mechanism | Ensures originality and quality, but adds publication delays |
| **Digital transformation & open access** | Human-digital | Democratization of knowledge | Accessibility, more rapid dissemination, information overload |
| **Rise of predatory journals** | Human-digital | Exploitative practices, forced to publish or perish | Erosion of trust in academic publications |
| **Global inequalities and biases in terms of race, gender and geography** | Human-digital | Positive discrimination, more diversity in research production | Equity challenges and power imbalances |
| **Social media & influencers** | Human-digital | Trend-setting and informal norms | Rapid dissemination, potential for misinformation, |
| **Generative Academia: co-intelligence in research and publication** | Human-AI | Democratization of research & content creation | Challenging academic integrity, credibility, authorial authenticity and evaluation processes |

While GPT models and generative AI improve research productivity, their integration into academia demands careful balancing to preserve human creativity and reflexivity or critical thinking (Geraldi et al. 2024, Lindebaum & Fleming 2024, Floridi 2020). Bottom-up movements, especially the content of academic influencers via social media, rapidly transform academic norms and practices, reflecting this dynamic. A blend of top-down rigor with bottom-up insights and a shift of focus from tools to real-world practices within the broader history of academic integrity is required to address AI's ethical and practical challenges.

## 4. METHODS
*Exploratory Part and Research Questions*
YouTube and TikTok are among the most influential global social media platforms due to the dominance of video content (Statista, 2024). Our initial test revealed that YouTube's longer formats provided more detailed, valuable content in the field studied. Like traditional academic writing tutorials, these well-designed resources offered step-by-step guidance. Based on the preliminary random testing of the videos and the current scientific and societal debates in generative AI, we formulated the following research questions for systematic qualitative research:

- RQ1. How do academic influencers' video tutorials guide audiences in using ChatGPT for the publication process in a pipeline?

- RQ2. How can tutorials establish a generative knowledge workflow to improve productivity and efficiency?
- RQ3. How do tutorial videos tackle ethical concerns and fine-tune AI-generated content values?

With these research questions, we explore how influencer content contributes to building a generative knowledge production pipeline while addressing ethical concerns, evaluating its potential to drive a paradigm shift in academic publishing.

*Sampling and Corpus*

The goal was to create a consistent database focused on the content produced by academic and scientific influencers, specifically targeting videos that share knowledge, experiences, and tips related to the effective and ethical use of ChatGPT. The sampling period was 2024 summer, selecting the most popular language option, English. The keyword-based research always started with "ChatGPT"; after that, the operator was "AND," to which we added the keywords from the research questions and the preselected videos from the random search: "literature review," "research," "research questions," and "academic writing." As the platform's automatic filter category, we focused only on videos between 4 to 20 minutes in length. This length allows more precise hits and represents a higher number of viewers.

We focused only on the most popular videos that provided step-by-step tutorials (excluding a simple list of instructions, opinion or promotion videos). We focused on content creators with at least 500 followers, confirming their status as influencers via meaningful audience impact.

With this sampling logic, 53 videos remained in the corpus, produced by individual scholars and PhD students alone, and reaching 5,331,000 viewers in total. Considering each video influencer and their content, we collected metadata, including: name, country, age, gender, field, affiliation, follower count, topics, keywords, and discussions on plagiarism and ethical questions. To build the corpus, we compiled the transcripts of all the videos into a single text-based document, including the titles and links of the corresponding YouTube videos. In total, the transcripts present 120,667 words. To ensure accuracy and avoid spelling errors, the transcript of each YouTube video was manually reviewed and corrected as needed. The selected corpus is appropriately sized, given that the videos were rigorously chosen and have been viewed by a substantial audience, ensuring relevance to the study.

*Generating a Code System and Coding for Analysis*

After manually testing video structures, we identified three primary components in the step-by-step guides: *input*, *process*, and *output*. These categories were subsequently applied to analyze the content. Considering the large number of videos and their spoken language style and high word count, we generated a potential coding system using ChatGPT (GPT-4o). The initial prompt requested: "Use this text to break down into 3 categories/codes (with further subcategories, max 5 to summarize): (1) input (of a text, automation), (2) process (ethics vs. plagiarism), and (3) output (to produce academic publications). Use bullet points for categories/subcategories."

Based on the results, we refined the generated categories and secondary codes (e.g., human interaction, task, data input, ethical considerations, AI contribution, and publication process) alongside subcodes (e.g., drafts, utilization, readability, and AI creativity). We prioritized the coding rules specificity, focusing on detailed, direct interactions with AI generated outputs. We ignored any content in the YouTube videos that lacked specific context or examples.

This coding system resulted in relevant, actionable insights, streamlining the analysis of video content for academic workflows. Using the prompt: "Based on the texts submitted so far, what are the most common codes in the Input, Process, and Output categories?" we established a standard coding system. Then, we manually cross-checked and validated the AI-generated codes. Exclusion criteria were reapplied, and illustrative examples were gathered manually through automation (GPT-4o): "Please find illustrative examples and interpretative text parts to the above-generated codes." We selected the most informative examples for the analysis.

The manual and automated analyses yielded almost identical results, confirming the effectiveness of human-machine co-intelligence. This coding method improved productivity and demonstrated seamless collaboration between humans and AI, aligning with the central theme of this study on human-machine interaction.

*Analysis*

Content analysis and a conceptual model were the mixed methods. For the content analysis, we systematically evaluated the frequency and context of themes and keywords in the qualitative data, coding the occurrence of terms related to academic integrity, generative AI/ChatGPT usage, and shifts in research practices or norms based on the research questions. Steps:

Step 1: Identifying key themes: generative AI-assisted process, ethical issues, plagiarism
Step 2: Coding the data: marking each instance where these themes or keywords appear
Step 3: Quantifying frequency: themes and keywords within the data
Step 4: Contextual analysis: which these terms are used in similar or different contexts

In developing the conceptual model, we structured the way in which generative AI-driven research and publication can be applied across various academic disciplines.

Step 1: Identifying the key phases in the AI-human co-working flow
Step 2: Coding the main- and subcategories of the studied phases
Step 3: Collecting the most representative instructions for main and subcategories

This mixed-method approach offers a structured framework for analyzing the broader impact of generative AI in academia, providing a robust basis for evaluating the potential emergence of a paradigm shift.

**5. RESULTS**

The analyzed influencers comprised included academic scholars, scientific writing instructors, and academic language tutors. Their video channels primarily originate from the Global North—particularly the United States, United Kingdom, Canada, and Australia—while India and Israel are also represented. They operate independently, representing their channels and small businesses instead of institutions. Half are senior academics (two from prestigious universities), while the rest are junior scholars, mostly PhD students. Their high viewership stems from their influencer status and engaging presentation, using filters, style shifts, and dynamic editing techniques.

Most influencers use the premium, thus paid version of ChatGPT, highlighting its advantages, which include advanced models, faster responses, PDF uploads, priority feature access, better text analysis, and improved research assistance. Commenters often express gratitude, noting how the videos address challenges in PhD work, research, and supervision. No negative feedback or criticism indicates that the video audience largely accepts the content. Despite targeted searches, no similar videos from institutions or research communities with comparable viewership were identified.

### Answering RQs

*RQ1: How do academic influencers' video tutorials guide audiences in using ChatGPT for the publication process in a pipeline?*

A structured human-AI workflow is defined in the tutorials, integrating ChatGPT into academic writing and publication. It divides the process into three key stages: *Input*, *Process, and Output*, each with corresponding elements (see Figure 1).

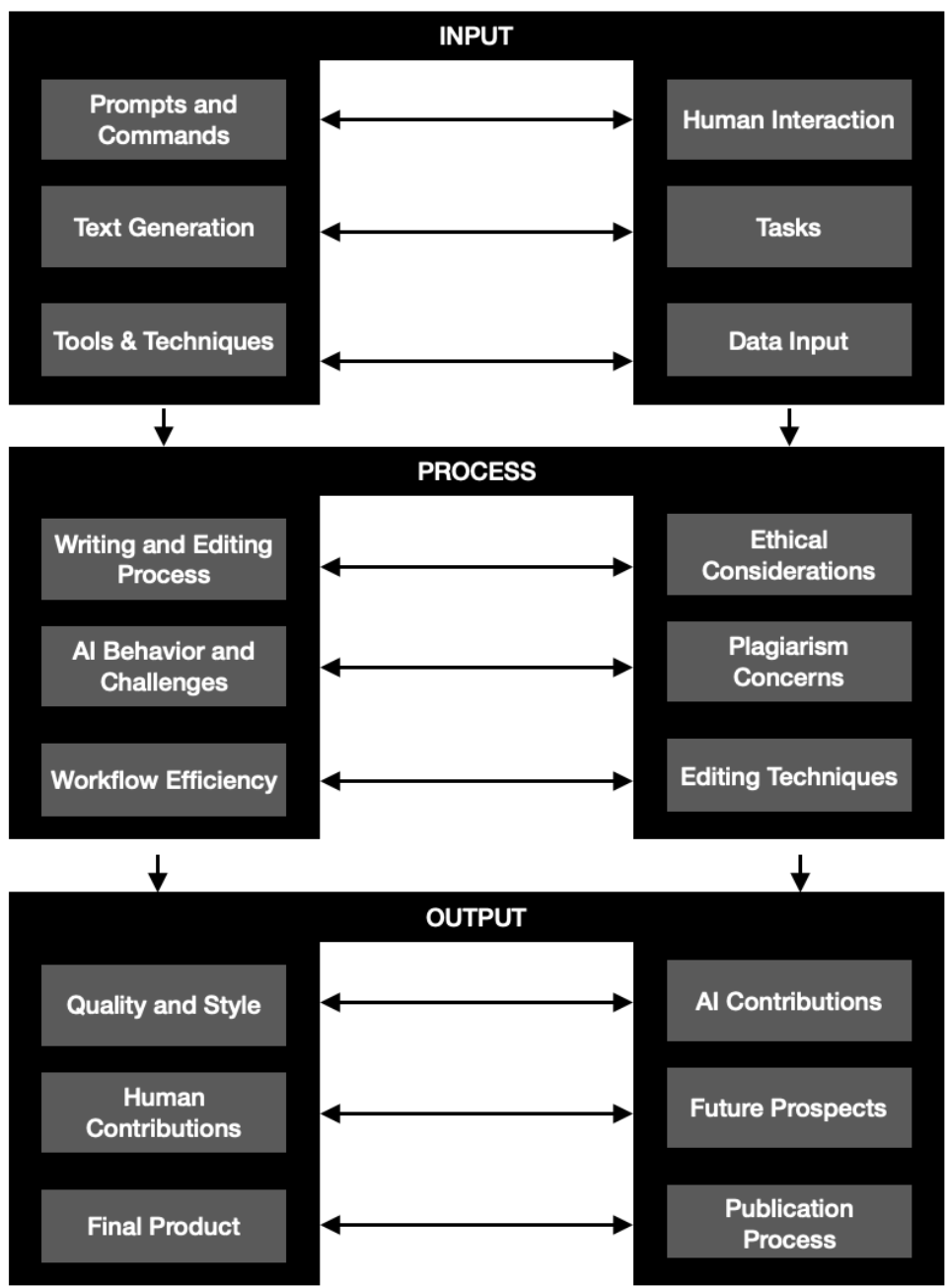

*Figure 1.*
*Generative Knowledge Production Pipeline. Source: The Authors*

The input phase includes prompts, text generation, and the tools and techniques used, which involve human interaction, tasks, and data input. The process phase focuses on writing, editing, AI behavior challenges, and workflow efficiency, balanced against ethical considerations, plagiarism concerns, and editing techniques. Finally, the output phase examines the quality of the final product, human contributions vs. AI's role, with an eye on the publication process. Academic influencers in the ChatGPT domain might suggest that this workflow is logical because it acknowledges the interplay between human and AI contributions at every stage, supporting academic integrity by incorporating checks for ethical considerations and plagiarism.

At first glance, this appears to be a well-prepared pipeline for automation, particularly when the goal is to understand and encode the human-AI co-intelligence working process in collaborative knowledge production. The structured and sequential approach effectively improves productivity and efficiency, making it a pipeline for AI-assisted research workflows. However, a few critical aspects remain underexplored despite its strengths in systematization and automation. In particular, quality control or feedback-based optimization requires deeper analysis to ensure accuracy, reliability, and ethical compliance in automated scholarly outputs. Without more rigorous refinement and validation mechanisms, automation risks reinforcing biases, inaccuracies, and ethical issues, potentially compromising academic integrity. These aspects are underrepresented in the tutorial videos. Thus, further pipeline sequences should be built for the automated human-AI co-working process for credibility and academic integrity.

*RQ2: How can tutorials establish a generative knowledge workflow to improve productivity and efficiency?*

To establish a generative knowledge production pipeline, the video tutorials emphasize the importance of specific and well-crafted prompts during the input phase as a critical factor in improving the efficiency, effectiveness, and productivity of using ChatGPT for academic work. By adjusting and refining prompts, users can improve the clarity and relevance of AI-generated content, leading to more precise and valuable outputs. According to most influencers, iterative feedback loops and sequential command strategies further boost productivity by effectively refining results and expanding content. The instructions are relatively simple and can be self-evident:

> "Sometimes we don't know where to start with ChatGPT and need to know the best prompt to put into it."

> "You might iteratively take your first prompt and realize, let's drill down to a little more."

With regard to the process phase, the influencers highlighted specific prompt-crafting, automation, and productivity tools. They argued that ChatGPT improves workflow by automating tasks such as drafting, editing, and proofreading, allowing users to focus on the more complex aspects of their work. According to their approach, this workflow leads to significant time savings and productivity boosts. They mentioned:

> "All these things that would have taken you hours or maybe even days can now take just several minutes or even shorter, like literally seconds, when you use ChatGPT."

> "ChatGPT is the best tool for researching on the internet, but only if you know the crucial steps to set it up correctly."

Analyzing the output phase, the influencers mainly highlighted improvements in writing quality, stylistic consistency, readability, and tone appropriateness. They also stressed the importance of ensuring that content is compelling, engaging, and meets the desired productivity and efficiency standards for academic writing. Although they interpret the tool as a simple, helpful service, a few are skeptical regarding the output. In their words:

> "First, we're going to look at how ChatGPT can help you edit grammar. This one is probably the simplest. You put in your text prompt 'edit this to be grammatically correct' and ChatGPT will fix your typos, misspellings, and grammar issues."

> "AI-generated writing may be grammatically correct and technically proficient, but it lacks the artistry of an individual translating their heart and soul into a story."

According to these results, a clear objective and tailored prompt refinement are essential foundations to improve an effective and productive generative AI pipeline, followed by streamlined improvements in the process phase and the application of academic writing skills. However, this approach lacks a systematic feedback loop for iterative optimization. Additionally, the role of critical evaluation and verification remains underexplored, raising concerns about bias, credibility, and academic integrity. For a truly effective and scalable pipeline, these elements should be integrated into a continuous improvement cycle, ensuring reliability and alignment with scholarly standards.

*RQ3: How do tutorial videos tackle ethical concerns and fine-tune AI-generated content values?*
In the input phase, specific prompts and clear instructions help ensure that the generated content maintains originality and adheres to academic standards in the tutorial videos. At the same time, the influencers emphasized how important it is to refine and rewrite the generated texts to avoid plagiarism and improve the quality of AI-assisted work. They are particularly focusing on further tools to check plagiarism or originality.

> "Customize the Content: Rewrite the text [to] make sure it's not plagiarized, fact-check it because a lot of times the information in ChatGPT is not factually accurate, and add your own input so the content has your voice and is unique."

> "Now, we have essentially rewritten this content that was generated [by] ChatGPT and made it unique. Remember to fact-check and add your own input as well. If you take this content and paste it into Originality.ai, click on scan and just like that we have zero percent AI and 100 percent original. The key is to take these extra steps of rewriting the content and not just copying and pasting from ChatGPT."

*Reaching the process phase, influencers highlight how crucial it is to maintain academic integrity. They underscore the need for clear instructions, ethical usage, transparency, originality checks, and proofreading to avoid plagiarism. The emphasis is on responsible AI usage that supports academic integrity, but with floating questions and opinions, such as:*

> There are open questions about [whether] you need to declare that ChatGPT helped edit the manuscript. I'm of the opinion that the answer to that is yes.

> *The final question is [whether] any of this [is] ethical? Like is this the same as just hiring someone to write your thesis? And I would say no.*

> *Some professors at large institutions suggest their students should try out all AI tools so they familiarize themselves with them because they're going to be used more and more in the future of work.*

Finally, the output phase concerns originality, ethical writing, and proper citation practices. They mostly discuss ethical issues; plagiarism only appears in half of the videos, and AI hallucination is only mentioned in a few videos. Several tutorials highlight the risks of relying on AI-generated content without proper human oversight, stressing the need for responsible usage to maintain the credibility of academic work. To this, they promote further tools, such as Zotero and Grammarly, or browser extensions. Others highlight the co-intelligence movement beyond the human capacity or intentions:

> *"What you do not want to do with ChatGPT is use it to replace your writing. This is very important. Many YouTube videos teaching you how to use ChatGPT mess up because they're teaching you to use ChatGPT as a replacement for your writing. That is not what you want to do, and that's not going to fly with Amazon. Do not use it to replace your writing; use it as a supplement, as a tool to help you with the writing."*

> *"I did change a lot while I was editing this. I think in places I maybe took the editing a little farther than I originally intended but there were times when it was easier for me to just write a few lines of transition or fix the dialogue myself instead of having ChatGPT generate something better."*

The results reveal how the input phase ensures originality and prevents plagiarism, the process phase upholds academic integrity, and the output phase refines ethical writing, originality, and reliable citation. These overlapping areas require structured quality control for automation and enhanced reliability in a pipeline design.

*Analysis of the Key Results:*
*An Academic Paradigm Shift via Generative Knowledge Production Pipeline*
The structured workflow in influencer videos emphasizes a proactive approach to academic integrity, focusing on originality, ethical compliance, and human responsibility across three key phases within AI-assisted research and writing as follows:

- Input Phase: Academic integrity begins with aligning prompts and tools under human oversight, ensuring originality and credibility while requiring scholars to refine inputs actively for reliable AI-supported knowledge production.
- Process Phase: Influencers emphasize ethical compliance and vigilance in monitoring AI-generated content, guiding scholars to uphold academic standards through editing techniques and critical reflection—but encouraging productivity and creativity in using generative tools.
- Output Phase: Human contributions and "soul" remain central, following the requirements of the top-down approaches. Scholars are responsible for credible outputs supported by AI to maintain integrity through co-intelligence processes. Thus, the outputs are not plagiarism if a human interacts with the co-intelligence throughout the entire process.

The structured workflow in influencer videos represents a paradigm shift in academia and academic integrity, democratizing research and redefining human-technology interaction. This bottom-up approach emphasizes originality, ethical compliance, and human oversight across the input, process, and output phases. Scholars are encouraged to refine inputs, creatively and critically assess AI-generated content, and take accountability for final outputs, ensuring co-intelligence between humans and AI throughout the workflow. Such workflows presuppose that generative AI is now significantly integral to academic work, with processes built around prompt decisions while also incorporating traditional academic integrity norms.

This shift challenges traditional authorship and evaluation processes by addressing ethical concerns such as algorithmic ghostwriting, transparency, and imbalances in academic trust. While some videos downplay these issues, none promote uncritical AI use, suggesting a proactive stance on academic integrity. Unlike rigid top-down policies, this model empowers scholars to integrate innovation with ethical practices, fostering adaptability and inclusivity.

In this way, the results represent a Generative Knowledge Production Pipeline, redefining how academic standards evolve alongside technological advancements. This pipeline modernizes academic practices while safeguarding credibility, providing a balanced approach to ethical scholarship in the evolving generative AI era.

These findings validate the paradigm shift via prompt-driven human-AI co-working, highlighting the need for pipeline improvements—including quality checks and optimization—to ensure a well-structured framework suitable for a proof of concept (PoC).

## 6. DISCUSSION

The Generative Knowledge Production Pipeline explores how academic influencers shape norms and practices in AI-assisted research, emphasizing ethical compliance, originality, and human oversight across structured workflows in academia. While previous research highlights the potential for AI tools such as ChatGPT to improve research productivity and publication quality (Singer, 2024; Mollick, 2024), it lacks detailed examinations of informal, bottom-up influences in most cases, especially regarding the impact of academic influencers on social media platforms such as YouTube. This study fills that gap by exploring how social media influencers shape academic norms, offering structured workflows and ethical guidance, particularly through the input, process, and output phases. These insights extend the discourse on how AI should be integrated into academia while maintaining originality, ethical compliance, and human oversight.

Unlike traditional top-down academic policies, the influencer-driven model democratizes research and content creation, fostering tacit consensus on AI use. This includes strategies for crafting precise prompts, iterative feedback loops, and creative applications of AI tools—all aimed at maintaining integrity while improving productivity. These informally driven yet impactful frameworks align with evolving generative AI policies and bridge the gap between institutional standards and practical applications. This study highlights the convergence of bottom-up practices with top-down guidelines, demonstrating how dominant Global North frameworks shape discourse from both directions.

The Generative AI Knowledge Production Pipeline is relevant as it systematizes human-AI collaboration, demonstrating how generative tools can support research workflows. Its structured approach improves productivity and efficiency with valuable and ethical fine-tuning, making it applicable beyond academia as well. However, critical refinements are needed, particularly in quality control and optimization to ensure accuracy and academic integrity. Strengthening these areas will transform the pipeline into a scalable and indispensable framework for future scholarly and professional knowledge production, ensuring credibility, originality and integrity in AI-assisted content generation across diverse domains.

Accordingly, the revealed pipeline for a PoC is precious to universities, research institutions, publishers, and policy-makers if they search for implications for a proactive approach in Generative AI co-intelligence workflow (see Table 2). Institutions can adopt structured workflows from influencer content to develop official guidelines and training programs, integrating productivity and ethics into AI-assisted research. Academic influencers could also become pivotal in promoting transparency and creating standardized tutorials. Social media platforms, currently underused in disseminating accessible academic resources, provide an avenue for expanding equitable knowledge-sharing networks.

*Table 2*
*Academic Policy Implications for Reactive and Proactive Approaches.* Source: The Authors

| STAGE | Reactive Implications | Proactive Implications |
|---|---|---|
| INPUT | Institutional academic generative ethics | Integrating bottom up approach, co-intelligence and collaborative AI-scholars, community driven joint academic-business models and academic ethics by design, institutional subscription to generative tools |
| PROCESS | Individual scholars/research teams & institution-based ethics | Dynamic, real-time research-publication-ethics frameworks adaptable to AI developments and AI feedback loop, misinformation detection tools |
| OUTPUT | Publication ethics following the outdated academic integrity | Evolved publication standards integrating AI ethics by design for creativity and credibility, jointly with the top and emerging publishers, AI-curated content, ethical AI gamification/entertainment and audits, top-down and bottom up consensus |

This framework also highlights challenges, such as the potential for market-driven dynamics to dilute the ethical contributions of academic influencers, or replace them with synthetic actors in the future. Additionally, emerging generative tools like Scite.ai and Scholarcy lack sufficient influencer-driven content, which limits their impact. Proactive policies are required to address these gaps and guide ethical AI use that aligns with evolving technological trends and ensures long-term academic integrity.

Building on previous works (Rasul et al. 2023; Livberber & Ayvaz, 2023; Lim, 2023), this study emphasizes the importance of human agency in creativity, judgment, and ethical decision-making. Social media influencers' guides and tutorials demonstrate that co-intelligence systems—integrating human oversight and AI capabilities—can advance academic practices, fostering collaboration and productivity.

The findings suggest that a Generative AI Knowledge Production Pipeline should be applied in co-intelligence workflows, considering the trends emerging in parallel, such as real-time feedback loops, open-source models, and gamified incentives for ethical research. Proactive approaches must anticipate the impact of AI-driven research and publication, address the spread of misinformation, ensure transparency, and align with academic integrity. With their ability to reach diverse audiences, social media influencers can champion responsible AI use and advocate for aligning generative AI tools with the broader goals of academic and social good. By embracing collaborative, community-driven approaches, academia can evolve to meet the challenges of a rapidly changing technological landscape while dynamically applying ethical and scholarly standards.

## 7. LIMITATIONS

This study, conducted exclusively in English, reflects perspectives that are dominated by the Global North, highlighting the need for research in diverse linguistic and cultural contexts. The analysis of 53 videos, collectively reaching nearly 80 million users, demonstrates significant influence on academic practices but excludes smaller-scale influencers with fewer than 500 followers. Future research can address these gaps, examining the role of smaller influencers and exploring how top-down and bottom-up generative AI guidelines converge globally to improve academic research and integrity.

## 8. FUNDING

This work was supported by the European Union's Horizon Europe Research and Innovation Programme – NGI Enrichers, Next Generation Internet Transatlantic Fellowship Programme [Grant number: 101070125] awarded to Katalin Feher.

## 9. ACKNOWLEDGMENT

We gratefully acknowledge Viktor Horvath, data analyst at Ludovika University of Public Service, for his valuable contributions during the data collection and cross-checking phase.